# Energy spectrum of a generalized Scarf potential using the Asymptotic Iteration Method and the Tridiagonal Representation Approach

S. A. Al-Buradah[a], H. Bahlouli[a,b], A. D. Alhaidari[b]

[a] *Physics Department, King Fahd University of Petroleum & Minerals, Dhahran 31261, Saudi Arabia*
[b] *Saudi Center for Theoretical Physics, P.O. Box 32741, Jeddah 21438, Saudi Arabia*

**Abstract**: The well-known trigonometric Scarf potential is generalized by adding a sinusoidal term and then treated using the Asymptotic Iteration Method (AIM) and the Tridiagonal Representation Approach (TRA). The energy spectrum of the associated bound states are computed. For the AIM, we have improved convergence of the quantization condition that terminates the iterations asymptotically. This is accomplished by looking for the range of initial values of the space variable in the terminating condition that produces stable results (plateau of convergence). We have shown that with increasing iteration this plateau of convergence grows up rapidly to an optimal iteration number and then shrinks slowly to a point. The value of this point (or points) may depend on the physical parameters. The numerical results have been compared favorably with those resulting from the TRA.

*We are honored to dedicate this work to Prof. Hashim A. Yamani on the occasion of his 70$^{th}$ birthday.*

## I. INTRODUCTION

Almost all known methods of solution of the wave equation in quantum mechanics do very well (to a varying degree of accuracy) in problems with exactly solvable potentials. However, the real test of these methods is when the potential is not exactly solvable. There exist various methods to obtain approximate solutions. For instance, time-independent perturbation theory [1], WKB approximation [2], finite-element method [3, 4], and numerical method [5]. In atomic and molecular physics, due to the many body nature of their problems, numerical solution of the Schrödinger equation frequently employs self-consistent field approximation whereas in nuclear physics the Born-Oppenheimer approximation is used. Moreover, from the early days of quantum mechanics, numerical methods were already developed [6] in order to overcome the limitations of the number of exactly solvable problems. Therefore, in cases where analytical solutions are difficult to find or not possible, numerical methods are necessary [7]. In the past and in recent years, many developments in the numerical solution of the Schrödinger equation have appeared. Some of these methods include Matrix Diagonalization Method, Spectral Method, Discrete Variable Method, Runge-Kutta methods [8, 9]. In our work, we compare the numerical advantages of two methods to obtain the energy eigenvalues: the Asymptotic Iteration Method (AIM) and the Tridiagonal Representation Approach (TRA).

The AIM has been developed [10] to solve second order linear differential equations. For most Schrödinger equations with exactly solvable potentials, the AIM was found to reproduce very well the exact energy spectrum and wave functions [11–14]. Additionally, it gave very good results in the case of non-exactly solvable potentials [15–18]. Moreover, for analytically solvable potentials, the AIM formulation resulted in energy eigenvalues identical to those derived by other analytical means [19]. On the other hand, the TRA [20] has been successfully used to obtain the wave function, bound states energies and scattering phase shift analytically for a class of potentials that is larger than the traditional class. The J-matrix method that inspired the TRA, can also give the bound states and resonance energies associated with different short-range potentials. In particular, the approach was used for the Morse potentials [21], the inverse Morse potentials [22], the tamed Yukawa potential [23], the generalized Yukawa potential [24], the Hulthen potential [25], the Hellmann potential [26] and exponential-cosine-screened Coulomb potential [27]. However, there exist a class of non-conventional potentials with discrete spectra and in our work, we seek the solution of one such non-conventional potential in the 1D Schrödinger equation, namely the generalized trigonometric Scarf potential. For comparison purposes, we plan to apply both AIM and TRA to this problem. Generally, for problems with bound states we would like to obtain solutions of the time-independent wave equation $H|\psi_n\rangle = E_n|\psi_n\rangle$, which is an eigenvalue problem, where $\{\psi_n\}$ is the eigenfunction, $H$ is the Hamiltonian and $\{E_n\}$ is the corresponding eigenvalues (energy spectrum) [28]. In the AIM, we start by converting the Schrödinger equation into a standard form suitable for the method [10] and then use the quantization condition to terminate the iterations asymptotically and obtain the energy spectrum and eigenstates. On the other hand, in the TRA we expand the wavefunction in the space spanned by a complete set of suitable square integrable discrete basis. The basis set is chosen to produce a tridiagonal matrix representation [29, 30] for the wave operator so that the matrix wave equation becomes a symmetric three-term recursion relation for the expansion coefficients of the wavefunction. The recursion relation is then solved analytically in terms of orthogonal polynomials in the energy. We start briefly by introducing both methods and then use them to calculate the energy spectrum of a generalized version of the trigonometric Scarf potential. Therefore, we consider the following one-dimensional Schrödinger equation in a finite segment of space with a new generalized potential (in the atomic units $\hbar = m = 1$):

$$\left[\frac{d^2}{dx^2} - 2V(x) + 2E\right]\psi(x) = 0, \tag{1-1}$$

where the potential is defined by

$$V(x) = \begin{cases} V_0 + \dfrac{V_+ - V_- \sin(\pi x/L)}{\cos^2(\pi x/L)} + V_1 \sin(\pi x/L), & -\dfrac{L}{2} \leq x \leq \dfrac{L}{2} \\ \infty, & \text{outside} \end{cases} \tag{1-2}$$

Without the $V_1$ term, this is just the trigonometric Scarf potential, which belongs to the conventional class of exactly solvable problems with the following energy spectrum

$$E_n = \frac{\lambda^2}{2}\left[n + \frac{1}{2} + \frac{1}{2}\sqrt{\frac{1}{4} + \frac{2W_+}{\lambda^2}} + \frac{1}{2}\sqrt{\frac{1}{4} + \frac{2W_-}{\lambda^2}}\right]^2 + V_0, \tag{1-3}$$

where $W_\pm = V_+ \pm V_-$. The special case where $V_0 = V_\pm = 0$, which leaves only the new component, is just the infinite square well with sinusoidal bottom that does not have an exact solution in the conventional formulation of quantum mechanics. However, using the TRA, the Authors in [31] were able to obtain an exact solution for this problem. In the following sections, we will compute the energy spectrum associated with the full potential (1-2) using the AIM and the TRA to enable us to make a reasonable comparison of both approaches.

## II. THE ASYMPTOTIC ITERATION METHOD

The AIM was established to obtain analytic and/or numerical solutions of linear second order ordinary differential equations [10] and was applied to a wide range of problems in quantum mechanics. The method considers the following standard form of the equation:

$$\psi'' = k_0(x)\psi' + S_0(x)\psi \qquad (2\text{-}1)$$

where the prime stands for the derivative with respect to $x$, $k_0(x)$ and $S_0(x)$ are $C_\infty[\Omega]$ functions (infinitely differentiable in a domain $\Omega$ of the complex plane) and not necessarily bounded but such that $k_0(x) \neq 0$. In practice, $k_0(x)$ and $S_0(x)$ have sufficiently many continuous derivatives. However, the time-independent one-dimensional Schrodinger equation (1-1) is not in the standard AIM form (2-1) and hence we need to make a transformation to bring it to (2-1). That is we need to introduce a first order derivative into the Schrodinger equation. In most applications the functions $k_0(x)$ and $S_0(x)$ are polynomials or rational functions. We differentiate Eq. (2-1) and iterate up to $(n+1)^{th}$ and $(n+2)^{th}$ derivatives. Due to the linearity of the right side of (2-1) in $\psi(x)$ and $\psi'(x)$ we can easily obtain

$$\psi^{n+1}(x) = k_{n-1}\psi'(x) + S_{n-1}(x)\psi(x), \qquad \psi^{n+2}(x) = k_n\psi'(x) + S_n(x)\psi(x) \qquad (2\text{-}2)$$

where $k_n(x)$ and $S_n(x)$ are defined recursively as follows

$$k_n(x) = k'_{n-1}(x) + S_{n-1}(x) + k_0(x)k_{n-1}(x), \qquad S_n(x) = S'_{n-1}(x) + S_0(x)k_{n-1}(x) \qquad (2\text{-}3)$$

The ratio of $(n+1)^{th}$ and $(n+2)^{th}$ derivatives of $\psi(x)$ gives

$$\psi^{n+2}/\psi^{n+1} = k_n\left[\psi' + (S_n/k_n)\psi\right]/k_{n-1}\left[\psi' + (S_{n-1}/k_{n-1})\psi\right].$$

Therefore, as per the asymptotic iteration technique and for adequately large $n$, we impose the termination condition:

$$\frac{S_n(x)}{k_n(x)} = \frac{S_{n-1}(x)}{k_{n-1}(x)} = \chi(x), \qquad (2\text{-}4)$$

Asymptotically, the terminating function $\chi(x)$ is independent of $n$. Therefore, once the functions $k_0(x)$ and $S_0(x)$ are determined, the sequences $k_n(x)$ and $S_n(x)$ can be computed using (2-3) and the energy spectrum (eigenvalues $E_n$) of the Schrodinger equation Eq. (1-1) are then obtained from the roots of the termination condition

$$\Delta_n(x, E) = k_n(x, E)S_{n-1}(x, E) - k_{n-1}(x, E)S_n(x, E) = 0, \qquad n = 1, 2, 3, \ldots \qquad (2\text{-}5)$$

where $n$ is the number of iterations and (2-5) is called the quantized equation. Numerically, the eigenvalues of the lowest $n+1$ energy levels are obtained by the requirement that $\Delta_n(x, E)$ becomes vanishingly small for as large number of iterations as possible to achieve the desired accuracy. In general, the AIM sequences $k_n(x)$ and $S_n(x)$, $n = 0, 1, 2, \ldots$ depends on $x$ and the (unknown) energy $E$. Thus, the energy eigenvalue $E_n$ that solves (2-8) depends generally on $x$. However, for analytically solvable potentials, the termination condition (2-5) gives an expression that depends just on $E$ (independent of $x$). In such situations, the energy eigenvalues are simply the zeros of $\Delta_n(E)=0$. Thus, the condition $\Delta_n(E)=0$ gives the energy spectrum of the exactly solvable potentials. On the other hand, for problems that are not exactly solvable the termination condition (2-5) produces for each iteration an expression that depends on both $x$ and $E$. Nevertheless, physically the eigenvalues should not depend on the space variable. Therefore, we need to find a mechanism that selects a natural or ideal point $x=x_0$ that gives stable and convergent results independent of this point for as large range of values of $x_0$ as possible. We refer to this range of values of $x_0$ as the "plateau of convergence" or "plateau of stability". Ideally, the solution should not depend on the choice of $x_0$ which means that the computation of the roots of $\Delta_n(x_0,E)=0$ should be free of the choice of $x_0$. Traditionally, researchers who use the AIM choose $x_0$ as either the position of the minimum value of the potential under consideration or as the location of the maximum of the ground state wavefunction as dictated by physical intuition. In this work, we present a more systematic method to choose a suitable value for $x_0$ from within the plateau of stability, which is defined as the range of values of $x_0$ where the calculated energy eigenvalue is stable against variations in $x_0$. For this purpose, we solve the quantization equation Eq (2-5) for certain value of $x$ and then compute the eigenvalues $E_m^n(x)$ where $n$ refers to the number of iterations and $m$ refers to the order of the energy eigenvalues. As an example, the ground state energy $E_0^n$ is calculated for different values of $x$ and we observe that within a given range of $x$, called "*the plateau of stability*", the calculated energy eigenvalue does not change with $x$ (within the chosen accuracy of calculation). However, for our current problem we found that with an increase in the iteration the plateau widen rapidly then shrinks slowly to a point. Now, if this point depends on the physical parameters then we need to give a physical interpretation to such natural selection. Otherwise, we must find some mathematical or numerical justification for this choice of $x_0$. The general solution of the $2^{nd}$ order differential equation Eq. (2-4) is:

$$\psi(x) = \exp\left[-\int_x \chi_n(y)dy\right]\left\{C_2 + C_1\int_x \exp\left[\int_y (k_0(\tilde{y})+2\chi_n(\tilde{y}))d\tilde{y}\right]dy\right\} \tag{2-6}$$

It should be noticed that the first factor of this wavefunction is a polynomial solution, which is convergent and physical, while the second factor is non-physical and divergent in general. Therefore, $C_1=0$ in Eq. (2-6) and the regular wavefunction is determined by the following equation.

$$\psi(x) = C_2 \exp\left[-\int_x \chi_n(y)dy\right] = C_2 \exp\left[-\int_x \frac{S_n(y)}{k_n(y)}dy\right] \tag{2-7}$$

## III. THE TRIDIAGONAL REPRESENTATION APPROACH

This approach aims at solving the wave equation, $H|\psi\rangle = \varepsilon|\psi\rangle$, where $H$ is the Hamiltonian and $E$ is the energy eigenvalue. The energy eigenvalue is either discrete (for bound states) or continuous (for scattering states) [4-10]. The TRA solves this eigenvalue equation algebraically [11, 12]. In this approach, the wavefunction is written as a series in terms of square-integrable basis set $\{\phi_m(x)\}_{n=0}^{\infty}$. Therefore, we start by expanding the wavefunction in a complete basis as

$$\psi_E(x) = \sum_m f_m(E)\phi_m(x), \tag{3-1}$$

where $\{f_m(E)\}_{m=0}^{\infty}$ are the coefficients of expansion and the basis is chosen such that we get a tridiagonal symmetric matrix for the wave operator $J = H - E$. The following is the general form of the square-integrable basis:

$$\phi_m(x) = A_m w(y) P_m(y), \tag{3-2}$$

where $y = y(x)$, $A_m$ is a normalization constant, $P_m(y)$ is a polynomial of degree $m$ in $y$ and $w(y)$ is in the form of the weight function associated with the polynomials that should vanish on the boundaries of configuration space $x$. Consequently, the matrix elements of the wave operator, $J$, are defined by

$$J_{n,m} = \langle\phi_n|(H-E)|\phi_m\rangle = \langle\phi_n|\left[-\frac{1}{2}\frac{d^2}{dx^2} + V(x) - E\right]|\phi_m\rangle = 0 \tag{3-3}$$

Thus, the eigenvalues of the original Schrodinger equation are obtained from the eigenvalues of this wave operator matrix. The transformation $x \to y(x)$ is chosen such that the space of the basis becomes compatible with the domain of the Hamiltonian $x$. With this coordinate transformation, the wave operator (3-3) becomes:

$$J_{n,m} = \frac{1}{2}\langle\phi_n|\left\{-(y')^2\frac{d^2}{dy^2} - y''\frac{d}{dy} + 2U(y)\right\}|\phi_m\rangle = 0 \tag{3-4}$$

where $U(y) = V(x(y)) - E$. The prime on $y$ stands for the derivative with respect to $x$. In this work, we choose $P_m(y)$ as the Jacobi polynomial where $y \in [-1,+1]$ and $\{\phi_m(y)\}_{n=0}^{\infty}$ is written in the following form:

$$\phi_m(y) = A_m(1-y)^{\alpha}(1+y)^{\beta} P_m^{(\mu,\nu)}(y), \tag{3-5}$$

where $A_m = \sqrt{\frac{2m+\mu+\nu+1}{2^{\mu+\nu+1}}\frac{\Gamma(m+1)\Gamma(m+\mu+\nu+1)}{\Gamma(m+\nu+1)\Gamma(m+\mu+1)}}$, $P_m^{(\mu,\nu)}(y)$ is the Jacobi polynomial of degree $m$ in $y$ and the real dimensionless parameters $\{\alpha,\beta,\mu,\nu\}$ are such that $\alpha,\beta \geq 0$ and $\mu,\nu > -1$. These parameters will be chosen later to support the tridiagonal requirement of the matrix wave operator (3-4). The integration measure becomes $\int_{x_-}^{x_+}....dx = \int_{-1}^{+1}...\frac{dy}{y'}$. Compatibility with the weight function of the Jacobi polynomial and dimensionality requires that $y' = \lambda(1-y)^a(1+y)^b$ with $a$ and $b$ being real parameters and $\lambda$ is a positive real parameter having the dimension of inverse length. In fact, the tridiagonal requirement of the wave operator matrix (3-4) also leads to the same form for $y'$. Using the differential equation of the Jacobi polynomial and its differential property, the wave operator matrix elements are obtained after some manipulation (shown in the Appendix) as

$$J_{m,n} = A_m A_n \int_{-1}^{1} (1-y)^{2\alpha+a-1} (1+y)^{2\beta+b-1} P_m^{\mu,\nu}(y) \Bigg\{ -n(n+\mu+\nu+1) - (2\alpha\beta + \alpha b + \beta a) + \alpha(\alpha+a-1)\frac{(1+y)}{(1-y)}$$

$$+\beta(\beta+b-1)\frac{(1-y)}{(1+y)} - n\left(y + \frac{\nu-\mu}{2n+\mu+\nu}\right)\left(\frac{\mu+1-a-2\alpha}{(1-y)} - \frac{\nu+1-b-2\beta}{(1+y)}\right) - \frac{U(y)(1-y^2)}{(y')^2} \Bigg\} P_n^{\mu,\nu}(y) dy \quad (3\text{-}6)$$

$$+ A_m A_n \frac{2(n+\mu)(n+\nu)}{2n+\mu+\nu} \int_{-1}^{1} (1-y)^{2\alpha+a-1} (1+y)^{2\beta+b-1} P_m^{\mu,\nu}(y) \left(\frac{\mu+1-a-2\alpha}{(1-y)} - \frac{\nu+1-b-2\beta}{(1+y)}\right) P_{n-1}^{\mu,\nu}(y) dy = 0$$

Requiring that this matrix be tridiagonal and symmetric (as shown in the next section) transforms the wave equation into the following three-term recursion relation of the expansion coefficients of the wavefunction:

$$J_{n,n} f_n + J_{n,n-1} f_{n-1} + J_{n,n+1} f_{n+1} = 0 \qquad (3\text{-}7)$$

where all the $J_{n,m}$'s are functions of energy and potential parameters. Solving this recursion relation in terms of orthogonal polynomials gives the complete set of expansion coefficients leading to the phase shift and/or energy spectrum in addition to the associated wavefunction.

## IV. RESULTS AND DISCUSSION

Starting with the TRA, we consider the coordinate transformation $y = \sin(\lambda x)$ where $-L/2 \leq x \leq +L/2$, $\lambda = \pi/L$ making $a = b = \frac{1}{2}$ and $y \in [-1, +1]$. According to the orthogonality formula and the recursion relation of the Jacobi polynomials (given in the Appendix), the wave operator $J_{m,n}$ given by Eq. (3-6) will have a symmetric tridiagonal matrix representation only in three cases as follows:

(1) $2\alpha = \mu + 1/2$, $2\beta = \nu + 1/2$, $\dfrac{U(y)}{\lambda^2} = A\dfrac{1+y}{1-y} + B\dfrac{1-y}{1+y} + py + q$ \hfill (4-1a)

(2) $2\alpha = \mu + 1/2$, $2\beta = \nu + 3/2$, $\dfrac{U(y)}{\lambda^2} = A\dfrac{1+y}{1-y} + p\dfrac{y}{(1+y)} + \dfrac{q}{(1+y)}$ \hfill (4-1b)

(3) $2\alpha = \mu + 3/2$, $2\beta = \nu + 1/2$, $\dfrac{U(y)}{\lambda^2} = B\dfrac{1-y}{1+y} + p\dfrac{y}{(1-y)} + \dfrac{q}{(1-y)}$ \hfill (4-1c)

where $A$, $B$, $p$, and $q$ are real parameters with $A = \alpha(\alpha - 1/2)$, $B = \beta(\beta - 1/2)$. We will just consider the first case (4-1a) since the other two cases are well treated in the literature, the potential function in terms of the original space variable then reads

$$V(x) = V_0 + \frac{V_+ - V_- \sin(\lambda x)}{\cos^2(\lambda x)} + V_1 \sin(\lambda x), \qquad (4\text{-}2)$$

where the basis parameters are obtained from the physical parameters $V_\pm$, $V_0$ and $V_1$ as

$\mu^2 = \dfrac{1}{4} + \dfrac{2}{\lambda^2}(V_+ - V_-)$, $\nu^2 = \dfrac{1}{4} + \dfrac{2}{\lambda^2}(V_+ + V_-)$, $p = 2V_1/\lambda^2$ and $q = \dfrac{V_+}{\lambda^2} + \dfrac{2V_0}{\lambda^2} - \dfrac{2E}{\lambda^2}$. Thus, reality

requires that $V_+ \geq -2(\lambda/4)^2$ and $V_- \geq 0$. Figure 1, shows this potential for a given set of physical parameters. The tridiagonal matrix elements of the wave operator $J_{m,n}$ read as follows:

$$A_m A_n \int_{-1}^{1} (1-y)^\mu (1+y)^\nu P_m^{\mu,\nu}(y) \left[ \left( n + \frac{\mu+\nu+1}{2} \right)^2 - \frac{1}{4}(\mu^2+\nu^2-1/2) + py + q \right] P_n^{\mu,\nu}(y) dy \quad (4\text{-}3)$$

Using the orthogonality formula and the recursion relation of the Jacobi polynomials we obtain the matrix elements of the wave operator $J_{m,n}$ as follows:

$$J_{n,m} = \left[ \left( n + \frac{\mu+\nu+1}{2} \right)^2 + u_o - \varepsilon \right] \delta_{n,m} + u_1 \langle n|y|m \rangle \quad (4\text{-}4)$$

where $\varepsilon = 2E/\lambda^2$ and $u_i = 2V_i/\lambda^2$. Using the evaluation of $\langle n|y|m \rangle$ given in the Appendix, the wave operator matrix elements takes the following form:

$$J_{n,m} = \left[ u_1 \frac{\nu^2 - \mu^2}{(2n+\mu+\nu)(2n+\mu+\nu+2)} + \left( n + \frac{\mu+\nu+1}{2} \right)^2 + u_0 - \varepsilon \right] \delta_{n,m}$$

$$+ u_1 \left[ \frac{2}{2n+\mu+\nu} \sqrt{\frac{n(n+\mu)(n+\nu)(n+\mu+\nu)}{(2n+\mu+\nu-1)(2n+\mu+\nu+1)}} \delta_{n,m+1} + \frac{2}{2n+\mu+\nu+2} \sqrt{\frac{(n+1)(n+\mu+1)(n+\nu+1)(n+\mu+\nu+1)}{(2n+\mu+\nu+1)(2n+\mu+\nu+3)}} \delta_{n,m-1} \right] \quad (4\text{-}5)$$

The diagonal representation requires $V_1 = 0$ giving $E_n = (1/2)\lambda^2 \left[ n + (\mu+\nu+1)/2 \right]^2 + V_0$, which is the well know energy spectrum of the trigonometric Scarf potential. The three-term recursion relation associated with (4-5) reads as follows

$$\varepsilon f_n(\varepsilon) = \left[ u_1 \frac{\nu^2 - \mu^2}{(2n+\mu+\nu)(2n+\mu+\nu+2)} + \left( n + \frac{\mu+\nu+1}{2} \right)^2 + u_0 \right] f_n(\varepsilon)$$

$$+ u_1 \left[ \frac{2}{2n+\mu+\nu} \sqrt{\frac{n(n+\mu)(n+\nu)(n+\mu+\nu)}{(2n+\mu+\nu-1)(2n+\mu+\nu+1)}} f_{n-1}(\varepsilon) \right. \quad (4\text{-}6)$$

$$\left. + \frac{2}{2n+\mu+\nu+2} \sqrt{\frac{(n+1)(n+\mu+1)(n+\nu+1)(n+\mu+\nu+1)}{(2n+\mu+\nu+1)(2n+\mu+\nu+3)}} f_{n+1}(\varepsilon) \right]$$

If we take the special case $V_+ = V_- = 0$ which leads to $\mu^2 = \nu^2 = 1/4$ then this results in the infinite potential well in one-dimension with sine bottom $V_1 \sin(\lambda x)$ which was treated in [31]. Under these restrictions $\mu = \nu = 1/2$ and the three-term recursion relation equation reads as follows

$$\varepsilon f_n(\varepsilon) = \left[ (n+1)^2 + 2V_0/\lambda^2 \right] f_n(\varepsilon) + (V_1/\lambda^2)\left[ f_{n-1}(\varepsilon) + f_{n+1}(\varepsilon) \right] \quad (4\text{-}7)$$

Solving the three-term recursion relation (4-6) gives the energy spectrum $\varepsilon_n$ for the potential (4-3) (i.e., eigenvalues of the Schrödinger equation (1-1) with this potential). The results are reported below and compared to the AIM.

For the AIM, the Schrödinger equation with the potential (4-3) reads:

$$\left\{ -\frac{d^2}{dx^2} + 2\left[ V_0 + \frac{V_+ - V_- \sin(\lambda x)}{\cos^2(\lambda x)} + V_1 \sin(\lambda x) \right] - 2E \right\} \psi(x) = 0 \qquad (4\text{-}8)$$

Now, to obtain the suitable AIM form of the equation (2-1) one should take into consideration the boundary condition that requires the wavefunction to vanish at $x = \pm L/2$. To enforce satisfaction of the boundary conditions, we rewrite the wave function as follows:

$$\psi(x) = g(x) f(x) = \cos(\pi x/L) f(x) \qquad (4\text{-}9)$$

Substituting Eq.(4-9) in Schrodinger equation (4-8) gives:

$$f''(x) = 2\lambda \frac{\sin(\lambda x)}{\cos(\lambda x)} f'(x) + \lambda^2 \left( u_0 + u_1 \sin(\lambda x) + \frac{u_+ - u_- \sin(\lambda x)}{\cos^2(\lambda x)} - \varepsilon + 1 \right) f(x) \qquad (4\text{-}10)$$

where $u_i = \frac{2V_i}{\lambda^2}$ and $\varepsilon = \frac{2E}{\lambda^2}$. To bring the independent variable domain to $[-1, +1]$ we use the following coordinate transformation

$$y = \sin(\pi x/L), \quad -1 \leq y \leq +1 \qquad (4\text{-}11)$$

Then equation (4-10) can be written in a compact form as follows:

$$f''(y) = \frac{3y}{(1-y^2)} f'(y) + \left[ \frac{u_0 + u_1 y - \varepsilon + 1}{(1-y^2)} + \frac{u_+ - u_- y}{(1-y^2)^2} \right] f(y) \qquad (4\text{-}12)$$

From mathematical point of view, the points $y = \pm 1$ are regular singular points of the differential equation (4-12). This Equation is now amenable to the AIM where the characteristic functions $k_0(y)$ and $S_0(y)$ are defined by

$$k_0(y) = \frac{3y}{(1-y^2)}, \quad S_0(y) = \frac{u_0 + u_1 y - \varepsilon + 1}{(1-y^2)} + \frac{u_+ - u_- y}{(1-y^2)^2} \qquad (4\text{-}13)$$

The functions $k_n(y)$ and $S_n(y)$ are computed using the iteration formula (2-3). These will contain regular singular point and due to higher order derivative, the presence of the last term in $S_0(y)$ will play a very destructive role for the convergence of the termination condition. Thus to improve the convergence of the termination condition we need to perform a transformation that eliminates this spurious term as demonstrated shortly below in Eq. (4-15). The energy spectrum (eigenvalues $E_m^n$) will be found using the roots of equation (2-5). In this work we consider two cases by just changing the potential parameters:

**CASE 1:** $V_- = V_0 = 0$:

The characteristic functions $k_0(y)$ and $S_0(y)$ are written as follows:

$$k_0(y) = \frac{3y}{(1-y^2)}, \quad S_0(y) = \frac{u_1 y - \varepsilon + 1}{(1-y^2)} + \frac{u_+}{(1-y^2)^2} \qquad (4\text{-}14)$$

For convergence purposes, we eliminate the spurious term in (4-14) by writing

$$f(y) = (1-y^2)^\alpha g(y) \tag{4-15}$$

giving the following AIM equation for $g(y)$

$$g''(y) = \frac{(3+4\alpha)y}{(1-y^2)} g'(y) + \frac{\omega + u_1 y + 4\alpha(\alpha+1)}{(1-y^2)} g(y), \tag{4-16}$$

where we set $\omega = 1-\varepsilon$ and $\alpha = \frac{1}{4}\left(-1+\sqrt{1+4u_+}\right)$. Using Eq. (2-3) and Eq. (2-5) the functions $k_n(y)$ and $S_n(y)$ are calculated as follows:

$$k_1(y) = \frac{3 + u_1 y(1-y^2) + 4\alpha(2+\alpha) + 12y^2(1+\alpha)^2 + (1-y^2)\omega}{(1-y^2)^2}$$

$$S_1(y) = \frac{u_1(1+4y^2(1+\alpha)) + 4\alpha y(5+4\alpha)(1+\alpha) + y(5+4\alpha)\omega}{(1-y^2)^2}$$

$$k_2(y) = \frac{2u_1(1-y^2)(1+4y^2(1+\alpha)) + y(5+4\alpha)(9+20\alpha+8\alpha^2+2y^2(6+10\alpha+4\alpha^2)) + 2y(5+4\alpha)(1-y^2)\omega}{(1-y^2)^3}$$

$$S_2(y) = \frac{1}{(1-y^2)^3}\Big[u_1^2 y^2(1-y^2) + [8+12\alpha+4\alpha^2 + y^2(27+36\alpha+12\alpha^2)](4\alpha+4\alpha^2+\omega)$$

$$+ u_1 y(15+20\alpha+8\alpha^2 + y^2(20+28\alpha+8\alpha^2)) + (1-y^2)[2(2\alpha(1+\alpha) + u_1 y)\omega - \omega^2]\Big]$$

and so on. From the above calculations and by induction we conclude that $S_{2n}(y)$ and $S_{2n+1}(y)$ are polynomials in $\varepsilon$ of degree $n+1$ while $k_{2n}(y)$ and $k_{2n-1}(y)$ are polynomials in $\varepsilon$ of degree $n$ in $\varepsilon$. The Energy spectrum (eigenvalues) will be calculated using the termination equation (2-5),

For $n = 0$: $\Delta_0(y,\varepsilon) = a_1(\varepsilon) + y = 0$,

where $a_1(\varepsilon) = \frac{1}{u_1}\left[4\alpha(\alpha+1) + 1 - \varepsilon\right]$. With an appropriate choice of value $y = y_0$ we will obtain the first eigenvalue $\varepsilon_0$.

For $n = 1$: $\Delta_1(y,\varepsilon) = a_2(\varepsilon) + 2a_1(\varepsilon)y + y^2 = 0$,

where $a_2(\varepsilon) = \frac{1}{u_1^2}\left[4 + 24\alpha + 52\alpha^2 + 48\alpha^3 + 16\alpha^4 - (5+12\alpha+8\alpha^2)\varepsilon + \varepsilon^2\right]$. Again, with the same appropriate value of $y = y_0$ we obtain $\varepsilon_0, \varepsilon_1$

For $n = 2$: $\Delta_2(y,\varepsilon) = a_3(\varepsilon) + 3a_2(\varepsilon)y + 3a_1(\varepsilon)y^2 + y^3 = 0$, where

$$u_1^3 a_3(\varepsilon) = 264\alpha + 772\alpha^2 + 36 - 2u_1^2 + 1152\alpha^3 + 928\alpha^4 + 384\alpha^5 + 64\alpha^6$$

$$- (49 + 192\alpha + 288\alpha^2 + 192\alpha^3 + 48\alpha^4)\varepsilon + (14 + 24\alpha + 12\alpha^2)\varepsilon^2 - \varepsilon^3$$

giving $\varepsilon_0, \varepsilon_1, \varepsilon_3$. And so on.

The termination conditions for this problem are related as

$$\frac{\partial \Delta_n(y,\varepsilon)}{\partial y} = (n+1)\Delta_{n-1}(y,\varepsilon)$$

which, after integration, gives the following general from

$$\Delta_n(y,\varepsilon) = \sum_{k=0}^{n} c_k(\varepsilon) y^{k+1} + a_{n+1}(\varepsilon), \tag{4-17}$$

where $c_n(\varepsilon)$ is a linear combination of $a_n(\varepsilon)$ with coefficients that can be drawn from the Pascal triangle. It is worth mentioning that the general structure of $\Delta_n(y,\varepsilon)$ still holds independent of the specific potential.

$$\Delta_n(y,\varepsilon) = \sum_{k=0}^{n} c_k(\varepsilon) y^{k+p} + a_{n+1}(\varepsilon) \tag{4-18}$$

where $p$ is an integer that is potential dependent. The coefficients $c_n(\varepsilon)$ are functions of energy and potential parameters, $a_{n+1}(\varepsilon)$ are polynomials of degree $n+1$ in $\varepsilon$. Now, it is obvious that the function $\Delta_n(y,\varepsilon)$ is a polynomial of degree $n+1$ in $\varepsilon$. Thus, at the $n^{th}$ iteration there are $n+1$ eigenvalues that depend generally on the choice of $y_0$. Therefore, this problem with the generalized Scarf potential given by Eq. (4-2) will not have an analytic solution in the conventional formulation since the termination condition (2-5) depends on both $y$ and $\varepsilon$. Therefore, one has to find the best possible starting value $y = y_0$ that stabilizes the process of computation of the energy spectrum. As mentioned above, the energy eigenvalues obtained by calculating the roots of $\Delta_n(y_0,\varepsilon) = 0$ should be independent of the choice of $y_0$. Researchers traditionally choose this either at the minimum value of the potential or at the maximum value of the ground state wavefunction. However, the best choice of the starting value is observed to be critical only for the stability of the process, as well as to the speed of convergence. By means of the iteration fomula (2-3) and the the termiation condition (2-5), the eigenvalues are then computed. Our calculation shows the effect of (i) different choice of initial values $y = y_0$ and (ii) the number of iterations on the accuracy and convergence of the eigenvalues. In Table 1, we have considered only the lowest two Eigen-energies while varying $y_0$ in the range $-1 \leq y_0 \leq +1$. It is obvious that increasing the iterations the process stabilizes and the values of $y_0$ that are very close to the singular points lead to strong oscillations and divergence. It is also very clear in this table that for the ground state and for a given accuracy, 4 iterations produce Eigen-energies that are stable for any initial value within the range $-0.1 \leq y_0 \leq +0.1$ (the plateau of convergence). Increasing the number of iterations to 5, the plateau increases and the Eigen-energies are stable for any value in the given range (away from the singular points). Continuing to 6 iterations and further, the plateau shrinks to a point, which in this problem it is zero. Similarly, for the first excited states the plateau of convergence has the same behavior that it grows rapidly with the number of iterations up to an optimal number of iterations then shrinks slowly to a point (or points) as shown by the highlighted region of Table 1. For a better display of precision and convergence, the eigenvalues of the ground state are computed within the range $-0.1 \leq y_0 \leq +0.1$ and for iterations ranging from 1 to 100. It is obvious for AIM that the accuracy increases by increasing the number of iterations. The results are shown in Table 2 by comparing the ground state obtained via TRA (Table 3) with that obtained using AIM for different iterations. From the above two Tables we observe that by choosing a certain accuracy the termination condition can be solved for any starting value at certain number of iterations. For this number of iterations the eigenvalues are stable against variation in y within a given range. After that optimal number of iterations, the plateau starts to shrink to a point which can be considered as the ideal starting value of the space variable to ensure a good accuracy and convergence of the process. Therefore, we predict that for this problem, the best choice of $y_0$ is $y_0 = 0$. In many published research, the initial value of the space variable is set at zero without explanation. In this work we tried to have a closer look at the termination condition and its behavior depicted by Eq. (4.17) and studied its plateau of convergence. The termination condition depends on the iteration formula (2.3) which relies on taking an increasing number of derivatives. Thus, for higher number of iterations the termination condition becomes increasingly sensitive to any variations in $y$.

On first look at (4-17) it seems that if we insist that the roots of $\Delta_n(y,\varepsilon) = 0$ be independent of $y$ then one obvious choice is to set $y = 0$. In such a case what is left are the roots of the energy polynomials $a_{n+1}(\varepsilon)$. Hence, the roots of these polynomials represent the eigen-energies of the problem. This assertion leads us to conjecture that $a_{n+1}(\varepsilon)$ is proportional to the $(n+1)^{th}$ order polynomial generated by the three-term recursion relation in the TRA. Therefore, the technique of plateau of stability described above could be considered as one of the finest methods for choosing the best starting value of the space variable for the AIM. Most of our computations in this work were done using the computation software Mathematica. In Tables 3, the lowest Eigen-energies using both AIM and TRA methods are reported for comparison purpose.

**CASE 2:** $V_0, V_\pm = 0$:

We consider now the case of the infinite potential well with sine bottom $V(x) = V_1 \sin(\pi x/L)$. In the TRA this corresponds to $\mu = \nu = 1/2$. Equation (4-12) is then written as follows:

$$f''(y) = \frac{3y}{(1-y^2)} f'(y) + \frac{u_1 y - \varepsilon + 1}{(1-y^2)} f(y) \qquad (4\text{-}18)$$

The same procedures as in the previous case will be used for the calculation of the functions $k_n(y)$ and $S_n(y)$. Again, for this case the stability of the energy eigenvalues due to variations in $y_0$ has a similar behavior. For comparison purpose, the AIM and TRA results are reported in Table 4 for different parameters of the potential. Due to the rapid convergence and stabilization of the process for this case, the results are displayed for 10 iterations with an accuracy of more that 10 decimal degits. There is a good agreement between the two methods regarding the accuracy.

## V. CONCLUSION

The Asymptotic Iteration and Tridiagonal Representation techniques were used to calculate the eigen-energies of the generalized trigonometric Scarf potential. The work brings up the importance of the boundary conditions and coordinates transformation on the terminating condition of the asymptotic iterations in the AIM sequences associated with the Schrodinger wave equation. We also noticed that the existence of essential singularities would slow the convergence of the AIM. A remarkable observation however, is that the use of plateau of stability in the implementation of the AIM plays an important role in the convergence, precision as well as the stability of the numerical algorithm. However, for a given accuracy the plateau of convergence widens rapidly with increased iterations up to an optimum number of iterations. After which the plateau shrinks slowly with increasing the number of iterations and converges to a single point which, in this problem, seem to be independent of the potential parameters. We also gave a hand-waving argument in support of the usual choice of space parameter y=0 in the termination condition of the AIM.

**ACKNOWLEDGMENTS**:
We are grateful to material support of both the Saudi Center for Theoretical Physics (SCTP) and King Fahd University of Petroleum and Minerals (KFUPM) during the progress of this work. This work is extracted partly from the MSc thesis of Mr. Al Buradah at the physics department of KFUPM (2017).

**APPENDIX**

The Jacobi polynomial $P_n^{(\mu,\nu)}(y)$, where $y \in [-1,+1]$ and $\mu > -1, \nu > -1$ satisfies the following differential equation, differential relation, there-term recursion relation and orthogonality:

$$(1-y^2)\frac{d^2 P_m^{(\mu,\nu)}(y)}{dy^2} - \left[(\mu+\nu+2)y + \mu - \nu\right]\frac{dP_m^{(\mu,\nu)}(y)}{dy} + m(m+\mu+\nu+1)P_m^{(\mu,\nu)}(y) = 0 \quad \text{(A-1)}$$

$$(1-y^2)\frac{dP_m^{(\mu,\nu)}(y)}{dy} = -m\left(y + \frac{\nu-\mu}{2m+\mu+\nu}\right)P_m^{(\mu,\nu)}(y) + 2\frac{(m+\mu)(n+\nu)}{2m+\mu+\nu}P_{m-1}^{(\mu,\nu)}(y) \quad \text{(A-2)}$$

$$yP_m^{(\mu,\nu)} = \frac{\nu^2 - \mu^2}{(2m+\mu+\nu)(2m+\mu+\nu+2)}P_m^{(\mu,\nu)} + \frac{2(m+\mu)(m+\nu)}{(2m+\mu+\nu)(2m+\mu+\nu+1)}P_{m-1}^{(\mu,\nu)}$$
$$+ \frac{2(m+1)(m+\mu+\nu+1)}{(2m+\mu+\nu+1)(2m+\mu+\nu+2)}P_{m+1}^{(\mu,\nu)} \quad \text{(A-3)}$$

$$\int_{-1}^{+1}(1-y)^{\mu}(1+y)^{\nu}P_n^{(\mu,\nu)}(y)P_m^{(\mu,\nu)}(y)dx = \frac{2^{\mu+\nu+1}}{(2n+\mu+\nu+1)}\frac{\Gamma(n+\mu+1)\Gamma(n+\nu+1)}{\Gamma(n+1)\Gamma(n+\mu+\nu+1)}\delta_{nm} \quad \text{(A-4)}$$

Using the above and definition $\langle y | n \rangle = A_n (1-y)^{\mu/2}(1+y)^{\nu/2} P_n^{(\mu,\nu)}(y)$, we obtain

$$\langle n|y|m\rangle = \frac{v^2-\mu^2}{(2n+\mu+v)(2n+\mu+v+2)}\delta_{n,m}$$

$$+\frac{2}{2n+\mu+v}\sqrt{\frac{n(n+\mu)(n+v)(n+\mu+v)}{(2n+\mu+v-1)(2n+\mu+v+1)}}\,\delta_{n,m+1} \quad \text{(A-5)}$$

$$+\frac{2}{2n+\mu+v+2}\sqrt{\frac{(n+1)(n+\mu+1)(n+v+1)(n+\mu+v+1)}{(2n+\mu+v+1)(2n+\mu+v+3)}}\,\delta_{n,m-1}$$

Using the distributive property of the first order derivative operator, we obtain the following action of the first and second order differential operator on the basis elements (3-5):

$$\frac{d\phi_n}{dy} = A_n(1-y)^\alpha(1+y)^\beta\left\{\frac{d}{dy}+\frac{\beta}{1+y}-\frac{\alpha}{1-y}\right\}P_n^{(\mu,v)}(y) \quad \text{(A-6)}$$

$$\frac{d^2\phi_n}{dy^2} = A_n(1-y)^\alpha(1+y)^\beta \times$$

$$\left\{\frac{d^2}{dy^2}+\left(\frac{2\beta}{(1+y)}-\frac{2\alpha}{(1-y)}\right)\frac{d}{dy}+\frac{\alpha(\alpha-1)}{(1-y)^2}+\frac{\beta(\beta-1)}{(1+y)^2}-\frac{2\alpha\beta}{(1-y)(1+y)}\right\}P_n^{(\mu,v)}(y) \quad \text{(A-7)}$$

Substituting these together with $y' = \lambda(1-y)^a(1+y)^b$ and $y'' = \lambda^2(1-y)^{2a}(1+y)^{2b}\left(\frac{b}{1+y}-\frac{a}{1-y}\right)$

in the matrix wave equation (3-4), we obtain the matrix elements (4-6).

**TABELS CAPTION:**

**Table 1:** The effect of different initial value $y_0$ and number of iterations on the convergence and accuracy of the eigenvalues. We took the potential parameters as: $V_1 = 1, V_+ = 0.25, V_0 = V_- = 0$, and $L = 1$.

**Table 2:** The effect of different initial values $y_0$ and number of iterations on the accuracy of the ground state energy for different number of iterations with the potential parameter: $V_1 = 1, V_+ = 0.25, V_0 = V_- = 0$, and $L = 1$.

**Table 3:** The lowest levels of the energy spectrum for CASE 1 with the potential parameters: $V_1 = 1, V_0 = V_- = 0, V_+ = 0.25$, and $L = 1$ for both methods AIM (10 iterations ) and TRA (matrix size $N = 10$). We took $y_0 = 0$.

**Table 4:** The lowest levels of the energy spectrum for CASE 2 with the potential parameters $V_0 = V_\pm = 0, V_1 = 1$ and $L = 1$ for both methods AIM (10 iterations) and TRA (N=10).

**FIGURE CAPTION:**

**Figure 1:** The generalized Scarf potential for the parameters $V_0 = V_1 = 1, V_+ = 0.25, V_- = 0.1$, and $L = 1$.

**TABLE 1**

| | Iterations | -0.9 | -0.7 | -0.5 | -0.3 | -0.1 | 0 | 0.1 | 0.3 | 0.5 | 0.7 | 0.9 |
|---|---|---|---|---|---|---|---|---|---|---|---|---|
| $\varepsilon_0$ | 1 | 1.11121596 | 1.10615724 | 1.10253753 | 1.10022861 | 1.09912022 | 1.09898585 | 1.09911675 | 1.10013464 | 1.10210036 | 1.10494881 | 1.10862200 |
| | 2 | 1.09528288 | 1.09546731 | 1.09559105 | 1.09567252 | 1.09572924 | 1.09575353 | 1.09577777 | 1.09583377 | 1.09591196 | 1.09602618 | 1.09618936 |
| | 3 | 1.09575954 | 1.09575561 | 1.09575330 | 1.09575206 | 1.09575153 | 1.09575148 | 1.09575155 | 1.09575212 | 1.09575338 | 1.09575565 | 1.09575938 |
| | 4 | 1.09575085 | 1.09575091 | 1.09575094 | 1.09575095 | 1.09575096 | 1.09575096 | 1.09575096 | 1.09575097 | 1.09575098 | 1.09575101 | 1.09575106 |
| | 5 | 1.09575096 | 1.09575096 | 1.09575096 | 1.09575096 | 1.09575096 | 1.09575095 | 1.09575096 | 1.09575096 | 1.09575096 | 1.09575096 | 1.09575096 |
| | 6 | 1.09575096 | 1.09575096 | 1.09575096 | 1.09575096 | 1.09575096 | 1.09575096 | 1.09575096 | 1.09575096 | 1.09575096 | 1.09575096 | 1.09575098 |
| | 7 | 1.09575126 | 1.09575096 | 1.09575096 | 1.09575096 | 1.09575096 | 1.09575096 | 1.09575096 | 1.09575096 | 1.09575096 | 1.09575096 | 1.09575091 |
| | 8 | 1.09576026 | 1.09575096 | 1.09575096 | 1.09575096 | 1.09575096 | 1.09575096 | 1.09575096 | 1.09575096 | 1.09575096 | 1.09575096 | 1.09575136 |
| | 9 | 1.09589152 | 1.09575096 | 1.09575096 | 1.09575096 | 1.09575096 | 1.09575096 | 1.09575096 | 1.09575096 | 1.09575096 | 1.09575096 | 1.09567190 |
| | 10 | 1.09669582 | 1.09575095 | 1.09575096 | 1.09575096 | 1.09575096 | 1.09575096 | 1.09575096 | 1.09575096 | 1.09575096 | 1.09575096 | 1.09485162 |
| | 15 | −48.4023558 | 1.09575655 | 1.09575096 | 1.09575096 | 1.09575096 | 1.09575096 | 1.09575096 | 1.09575096 | 1.09575096 | 1.09575684 | −50.91596 |
| | 20 | −121.23852 | 1.15320760 | 1.09575021 | 1.09575096 | 1.09575096 | 1.09575096 | 1.09575096 | 1.09575096 | 1.09575147 | 17.129705 | −760.341065 |
| | 30 | −87475.723 | −206.58849 | 1.118114043 | 1.09575097 | 1.09575096 | 1.09575096 | 1.09575096 | 1.09575097 | 1.09075267 | −173.51422 | −79170.112 |
| | 50 | −29222.707 | −18517.54 | −503.87627 | 1.08537181 | 1.09575096 | 1.09575096 | 1.09575096 | 1.08153927 | 0..0000 | −17002.7 | −26036.89 |
| | 100 | −1.896722 | −25922.91 | −86230.156 | 0.00000 | 1.09575096 | 1.09575096 | 1.09575103 | 0.0000 | −86223.88 | −26531.56 | −1.89672 |
| $\varepsilon_1$ | 1 | 3.818650 | 3.904766 | 3.989442 | 4.072808 | 4.154974 | 4.195636 | 4.236034 | 4.316073 | 4.395164 | 4.473372 | 4.550756 |
| | 2 | 4.223949 | 4.214475 | 4.207675 | 4.203303 | 4.201145 | 4.200837 | 4.201014 | ,4.202745 | 4.206192 | 4.211224 | 4.217724 |
| | 3 | 4.196199 | 4.196484 | 4.196666 | 4.196777 | 4.196846 | 4.196873 | 4.196810 | 4.196964 | 4.197063 | 4.197220 | 4.197454 |
| | 4 | 4.196883 | 4.196879 | 4.196876 | 4.196874 | 4.196873 | 4.196873 | 4.196874 | 4.196874 | 4.196876 | 4.196878 | 4.196882 |
| | 5 | 4.196873 | 4.196873 | 4.196873 | 4.196873 | 4.196873 | 4.196873 | 4.196873 | 4.196873 | 4.196873 | 4.196873 | 4.196873 |
| | 6 | 4.196873 | 4.196873 | 4.196873 | 4.196873 | 4.196873 | 4.196873 | 4.196873 | 4.196873 | 4.196873 | 4.196873 | 4.196873 |
| | 7 | 4.196872 | 4.196873 | 4.196873 | 4.196873 | 4.196873 | 4.196873 | 4.196873 | 4.196873 | 4.196873 | 4.196873 | 4.196871 |
| | 8 | 4.196793 | 4.196873 | 4.196873 | 4.196873 | 4.196873 | 4.196873 | 4.196873 | 4.196873 | 4.196873 | 4.196873 | 4.196916 |
| | 9 | 4.197206 | 4.196873 | 4.196873 | 4.196873 | 4.196873 | 4.196873 | 4.196873 | 4.196873 | 4.196873 | 4.196873 | 4.200175 |
| | 10 | 4.193668 | 4.196873 | 4.196873 | 4.196873 | 4.196873 | 4.196873 | 4.196873 | 4.196873 | 4.196873 | 4.196873 | 4.208975 |
| | 15 | −19.8271 | 4.196697 | 4.196873 | 4.196873 | 4.196873 | 4.196873 | 4.196873 | 4.196873 | 4.196873 | 4.196795 | 0.000000 |
| | 20 | −17.7743 | 2.039172 | 4.196878 | 4.196873 | 4.196873 | 4.196873 | 4.196873 | 4.196873 | 4.196862 | 99.3084 | −361.064 |
| | 30 | −10370.12 | −21.6482 | 3.970720 | 4.196873 | 4.196873 | 4.196873 | 4.196873 | 4.196873 | 3.642828 | 25.3704 | −20192.02 |
| | 50 | −12256 | −8848.12 | −46.6361 | 4.272074 | 4.196873 | 4.196873 | 4.196873 | 4.166646 | 4.058064 | −11492.6 | −9821.21 |
| | 100 | −3.2202 | −2308.36 | −56002.4 | 0.00000 | 4.196874 | 4.196873 | 4.196873 | 0.00000 | −56004.09 | −2178.59 | −3.307368 |

**TABLE 2**

| Iterations \ y₀ | -0.1 | 0 | 0.1 |
|---|---|---|---|
| 1 | 1.0991202248315695 | 1.09898585279422 | 1.0991167526968348 |
| 2 | 1.0957292393803957 | 1.095753527922459 | 1.0957777701723037 |
| 3 | 1.0957515331417946 | 1.095751480043435 | 1.095751554739332 |
| 4 | 1.095750956836507 | 1.095750959070117 | 1.0957509613421887 |
| 5 | 1.0957509589138086 | 1.095750958910591 | 1.09575095891494 |
| 6 | 1.0957509588952719 | 1.095750958895319 | 1.09575095889537 |
| 7 | 1.095750958895318 | 1.095750958895318 | 1.0957509588953176 |
| 8 | 1.0957509588953167 | 1.095750958895317 | 1.095750958895317 |
| 9 | 1.095750958895314 | 1.095750958895318 | 1.0957509588953156 |
| 10 | 1.0957509588953172 | 1.095750958895317 | 1.095750958895318 |
| 15 | 1.0957509588953178 | 1.095750958895317 | 1.0957509588953191 |
| 20 | 1.0957509588953054 | 1.095750958895317 | 1.0957509588953052 |
| 30 | 1.0957509588953414 | 1.095750958895317 | 1.0957509588954089 |
| 50 | 1.0957509588941203 | 1.095750958895318 | 1.0957509588954446 |
| 100 | 1.0957509556475724 | 1.095750958895317 | 1.0957510397379888 |

**TABLE 3**

| $\varepsilon_n$ | AIM | TRA |
|---|---|---|
| 0 | 1.095750958895317 | 1.095750958895317 |
| 1 | 4.196873325806087 | 4.196873325806094 |
| 2 | 9.292844177187838 | 9.29284417718781 |
| 3 | 16.389252331506132 | 16.38925233150634 |
| 4 | 25.485790081950977 | 25.48579008195058 |
| 5 | 36.58237913869347 | 36.58237913866615 |
| 6 | 49.678992506294904 | 49.67899250636078 |
| 7 | 64.77561899613518 | 64.77561887661372 |
| 8 | 81.87225251919959 | 81.8727900682211 |
| 9 | 100.97373751566363 | 100.9688405503092 |

**TABLE 4**

| $\varepsilon_n$ | AIM $y_0 = -0.1$ | AIM $y_0 = 0$ | AIM $y_0 = 0.1$ | TRA |
|---|---|---|---|---|
| $\varepsilon_0$ | 0.9965804414948887 | 0.9965804414948881 | 0.9965804414948877 | 0.9965804414948881 |
| $\varepsilon_1$ | 4.001366326993638 | 4.0013663269936615 | 4.001366326993682 | 4.0013663269936615 |
| $\varepsilon_2$ | 9.000586656216685 | 9.000586656216424 | 9.000586656216043 | 9.000586656216425 |
| $\varepsilon_3$ | 16.00032590879089 | 16.000325908792934 | 16.000325908795052 | 16.000325908792927 |
| $\varepsilon_4$ | 25.000207394742418 | 25.000207394728182 | 25.000207394715762 | 25.000207394728175 |
| $\varepsilon_5$ | 36.00014358054208 | 36.00014358054704 | 36.00014358054918 | 36.0001435805457 |
| $\varepsilon_6$ | 49.00010529186438 | 49.00010529226843 | 49.000105292787566 | 49.000105292271191 |
| $\varepsilon_7$ | 64.00008070727844 | 64.00008065049936 | 64.00008070322694 | 64.00008052770214 |
| $\varepsilon_8$ | 81.00003408297702 | 81.00006328059685 | 81.00009189292832 | 81.00060387125417 |
| $\varepsilon_9$ | 100.00613182329307 | 100.00494124743338 | 100.0059053176835 | 100.000000000000 |

**FIGURE 1**

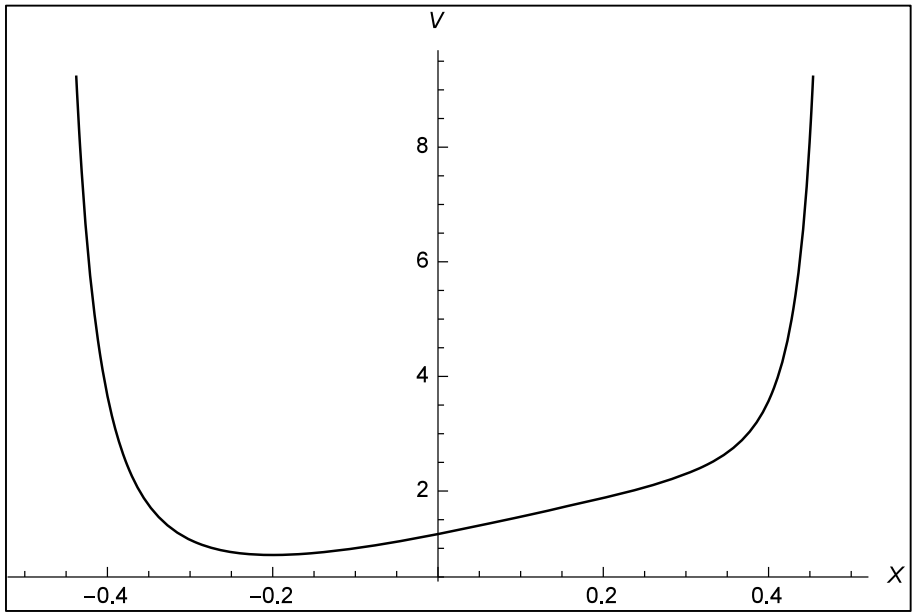